\documentclass{aa}

\usepackage{newtxtext,newtxmath}
\usepackage{graphicx}
\usepackage{longtable}

\usepackage[T1]{fontenc}
\usepackage{ae,aecompl}

\usepackage{graphicx}	
\usepackage{amsmath}	
\usepackage{amssymb}	

\begin{document}

   \title{Revisiting the 16 Cygni planet host at unprecedented precision and exploring automated tools for precise abundances}

   \author{M. Tucci Maia \inst{\ref{udp},\ref{lna}}
   \and J. Mel\'{e}ndez \inst{\ref{usp}}
   \and D. Lorenzo-Oliveira \inst{\ref{usp}}
   \and L. Spina \inst{\ref{usp}}
   \and P. Jofr\'{e} \inst{\ref{udp}}
   }

\authorrunning{Tucci Maia et al.}
\titlerunning{Revisiting the 16 Cygni}
\offprints{ \\ 
M. Tucci Maia, \email{marcelo.tucci@mail.udp.cl}
}

   \institute{N\'ucleo de Astronom\'ia, Facultad de Ingenier\'ia, Universidad Diego Portales,  Av. Ej\'ercito 441, Santiago, Chile \label{udp}
         \and Laboratorio Nacional de Astrof\'isica, Rua dos Estados Unidos 154, Itajub\'a, 37504-364 Minas Gerais, MG, Brazil \label{lna}
         \and{Universidade de S\~ao Paulo, Departamento de Astronomia do IAG/USP, SP, Brazil.\\}\label{usp}
         }

   \date{Received XXXX, 2019; accepted XXX, 2019}

  \abstract
   {The binary system 16 Cygni is key in studies of the planet-star chemical composition connection, as only one of the stars is known to host a planet. This allows us to better assess the possible influence of planet interactions on the chemical composition of stars that are born from the same cloud and thus, should have a similar abundance pattern. In our previous work, we found clear abundance differences for elements with Z$\leq30$ between both components of this system, and a trend of these abundances as a function of the condensation temperature (T$_{c}$), which suggests a spectral chemical signature related to planet formation. In this work we show that our previous findings are still consistent even if we include more species, like the volatile N and neutron capture elements (Z $>$ 30).
   We report a slope with T$_{c}$ of $1.56 \pm 0.24 \times 10^{-5}$ dex K$^{-1}$, that is good agreement with both our previous work and recent results by Nissen and collaborators. We also performed some tests using ARES and iSpec to automatic measure the equivalent width and found T$_c$ slopes in reasonable agreement with our results as well.
   In addition, we determine abundances for Li and Be by spectral synthesis, finding that 16 Cyg A is richer not only in Li but also in Be, when compared to its companion. This may be evidence of planet engulfment, indicating that the T$_{c}$ trend found in this binary system may be a chemical signature of planet accretion in the A component, rather than a imprint of the giant planet rocky core formation on 16 Cyg B.
   }
\keywords{Sun:abundances --
                stars:abundances --
                stars:solar-type --
                planet-star interactions
               }

\maketitle

%




\section{Introduction}

The most accepted hypothesis of star formation is the nebular collapse, when stars are formed from gravitationally unstable molecular clouds. Therefore, it is expected that stars that are born from the same interstellar material cloud should have the same abundance pattern during the Main Sequence, with exception of the light elements Li and Be, that can be destroyed in regions deeper than the convective zone, in solar type stars. Thereby, differences in the chemical content of stars born from the same natal cloud, may suggest that extra processes, that are not necessary connected to stellar evolution, may have influenced its photospheric chemical composition. In particular, planet formation or planet engulfment may imprint important chemical signatures in the host star. 

Such phenomenon is expected to leave subtle signs on the stellar abundance pattern, of the order of 0.01 dex \citep{cha10}, that can only be detected with a high-precision analysis.
This can only be achieved with the differential method \citep{nis18}, that requires a comparison between the target star with a similar star of known parameters, that will serve as the standard for the abundance calculations. So, the sample should be restricted to objects that are very similar among themselves, like in the case of solar twins\footnote{More recently, solar twins have been defined as stars with effective temperature within 100K; log g and [Fe/H] within 0.1 dex from the Sun \citep{ram14}.}. 

Following this premise, \cite{mel09} analyzed the abundances of 11 solar twins, achieving a high-precision abundance determination, with uncertainties of $\sim$ 0.01 dex, and found not only a depletion of refractory elements, when compared to the average of the sample, but also a trend with condensation temperature (T$_{c}$). The authors suggested that the correlation of the refractory elements abundances with condensation temperature is probably due to rocky planet formation. This hypothesis has been corroborated by \cite{cha10}, who showed that the depleted material in the Sun's convective zone is comparable to the mass of terrestrial planets of our Solar System (see also \cite{gal16}).

However, other hypothesis have been proposed to explain the abundance trend, like the stellar environment in which the star was formed \citep{one14}, although according to the recent theoretical estimates by \cite{gus18}, the mechanism is hardly significant; dust segregation in the protostellar disc \citep{gai15}; the influence of the stellar age \citep{adi14}; and the planet engulfment scenario \citep{spi15}.

In this context, twin stars in binary systems are extremely important, because the effects connected to the stellar environment of its formation and to the Galaxy chemical evolution, would be canceled out in a comparative analysis between both components. Thus, investigating wide binaries can bring more light into the subject of planets interactions (or other astrophysical event) influencing photospheric abundance of their host stars.

Some authors have already reported a T$_{c}$ trend on binary stars. \cite{tes16} found an abundance trend on the WASP94 system, where both stars are planet hosts. The planet-hosting binaries XO-2N/XO-2S \citep{ram15,bia15,tes15}, HD 133131A/B \citep{tes16b} and HAT-P-4 \citep{saf17}, also show chemical anomalies most likely due to planets, and the binary $\zeta^{1,2}$ Ret, where one of the stars hosts a debris disk, also shows a trend with condensation temperature \citep{saf16, adi16b}. Albeit no differences have been found in the HAT-P-1 \citep{liu14}, HD80606/HD80607 \citep{saf15,mac16} and HD20782/HD20781 \citep{mac14} and HD 106515 \citep{saf19}, the evidences are inconclusive in the latter three due to the high abundance errors. Indeed, a more precise abundance analysis of the pair HD80606/HD80607 by \cite{liu18}, shows small but detectable abundance differences between the binary components. A binary system of twin stars with large abundance differences, is HD 240429/HD 240430 \citep{oh18}, for which no planets are known yet. 
Furthermore, it was found that Kepler-10, a star hosting a rocky planet, is deficient in refractory elements when compared to stars with similar stellar parameters and from the same stellar population \citep{liu16}.

For 16 Cygni, a binary pair of solar twins, where the B component hosts a giant planet \citep{coc97}, \cite{law01} clearly detected that 16 Cyg A is more metal rich $\Delta$[Fe/H] = +0.025$\pm$0.009 dex than its companion.
Later, \cite{ram11} expanded the analysis to 23 chemical elements, showing abundance differences in all of them by about +0.04 dex and finding a T${_C}$ trend similar to \cite{mel09}, when the binary stars are compared to the Sun. This was confirmed in our previous work \citep{tuc14}, where we show that 16 Cyg A is $0.047 \pm 0.005$ dex metal richer than B and also finding a T$_{c}$ slope of $+1.99 \pm 0.79 \times 10^{-5}$ dex K$^{-1}$ for the refractory elements, as reported in \cite{ram11}). This result was then associated with the rocky core formation of the gas giant 16 Cyg Bb. Recently, \cite{nis17} also found a $\Delta$[Fe/H](A-B)= $+0.031 \pm 0.010$ dex and a T$_{c}$ slope of $+0.98 \pm 0.35 \times 10^{-5}$ dex K$^{-1}$.

In contrast, there are studies that challenge the  metallicity difference and the T$_c$ trend between the two components of this system.
\cite{sch11} find a T$_{c}$ trend for both stars relative to the Sun but, however, do not find any significant abundance differences between the pair, in agreement with \cite{del00} and \cite{tak11}.
Also, \cite{adi16b}, analyzing the case of $\zeta^{1,2}$ Ret, argues that the T$_c$ slope trend could be due to nonphysical factors and related to the quality of spectra employed, which is expected as high precision abundances can only be obtained in spectra of adequate quality.

In this context, the initial motivation for this work is to assess if, by revisiting this binary system now with better data with higher resolving power, higher S/N and broader spectral coverage, our previous results (obtained with lower resolving power) would still be consistent; in addition to provide improvements in the precision of the abundance determination, we include the analysis of elements that were not available before. We also challenge our results by employing automated tools to derive stellar parameters and T$_c$ while using the same methodology in all of the cases.

On the following sections, we will present the differential abundances of 34 elements and also the abundances of Li and Be, through spectral synthesis, which may present a possible evidence of planetary engulfment on 16 Cyg A.

\section{Data and analysis}

\subsection{Observations and data reduction}

The observations of 16 Cyg A and B were carried out with the High Dispersion Spectrograph \citep[HDS;][]{nog02} on the 8.2m Subaru Telescope of the National Astronomical Observatory of Japan (NAOJ), located at the Mauna Kea summit, in June 2015. Besides the 16 Cyg binary system, we also observed the asteroid Vesta, which was used as an initial reference for our differential analysis.

For the optical, we obtained an S/N ratio of $\sim$ 750 at 600nm and $\sim$ 1000 at 670nm (Li region), on the highest resolution possible (R$\sim$160 000) using the 0.2" slit.
The UV observations with HDS were made using the 0.4" slit, which provides a R = 90 000 that results in an S/N$\sim$350 per pixel at 340 nm, corresponding to the NH region, and S/N$\sim$200 at 310 nm (Be region). This gave us the opportunity to analyze volatile elements like nitrogen and neutron-capture elements in the UV with S/N $>$ 300. 

The stars from the binary system and the Sun (Vesta) were both observed using the same instrumental setup to minimize errors in a differential analysis, which requires comparisons between the spectra of all sample stars for the continuum placement and comparison of the line profiles, to achieve consistent equivalent width (EW) measurements.

The extraction of the orders and wavelength calibration were performed immediately after the observations by Subaru staff, with routines available at the observatory. The continuum normalization and Doppler correction were performed using standard routines with IRAF.

\subsection{Stellar parameters}

\begin{figure}
\centering
\includegraphics[width=1.0\columnwidth]{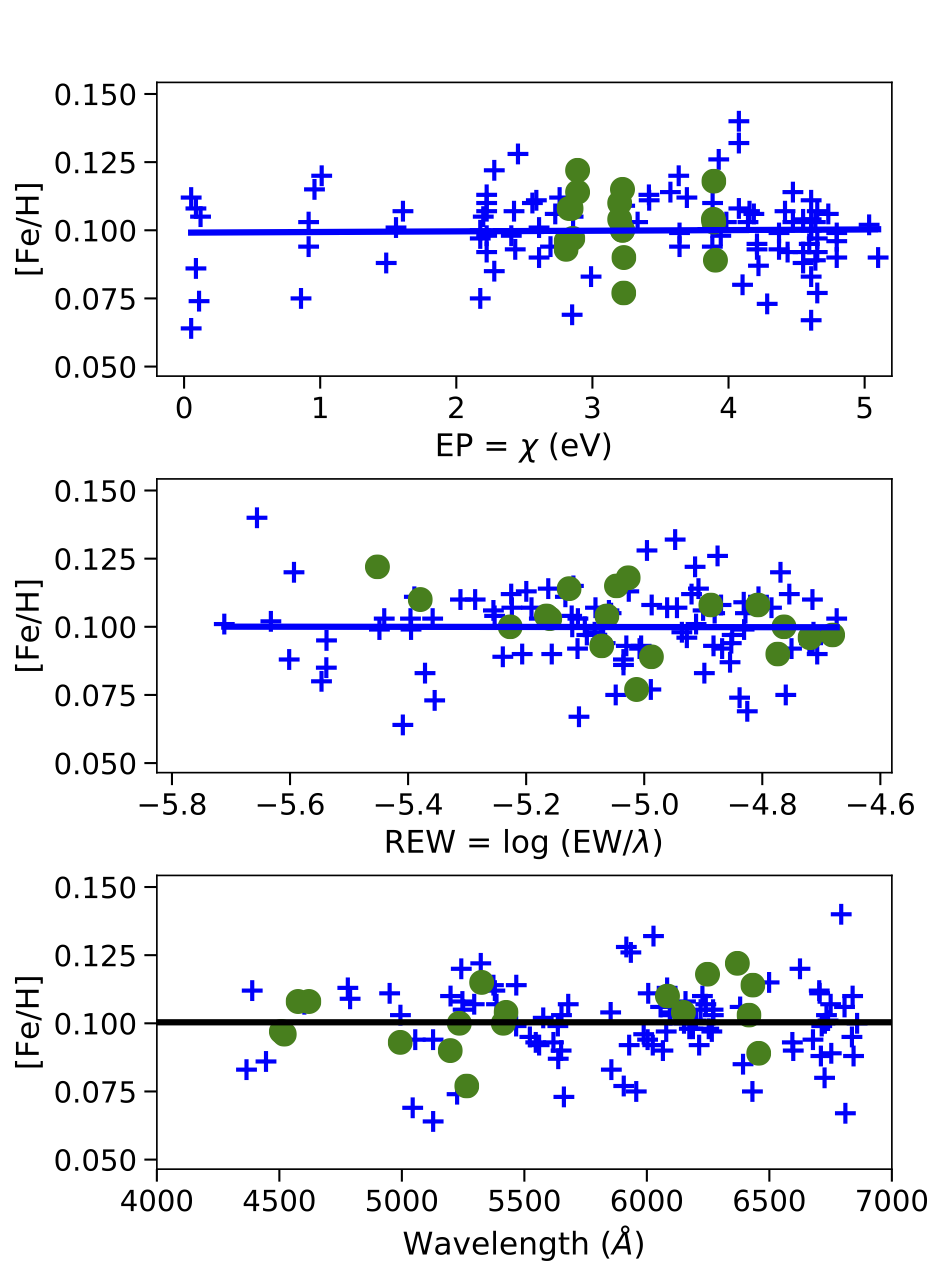}
\caption{Excitation and ionization equilibrium of Fe abundances (manually measured) using the Sun as the standard star for 16 Cyg A. Crosses represent Fe I and filled circles Fe II.}
\label{iso_sun}
\end{figure}

Our method to determine stellar parameters and elemental abundances follows the approach described on previous papers \citep[e.g.;][]{ram11,ram14,mel09,mel12,tuc16,spi16},
by imposing differential excitation and ionization equilibrium for Fe I and Fe II lines (Figure \ref{iso_sun}). Since the 16 Cygni system is a pair of solar twins, they have similar physical characteristics to the Sun, and thus we initially used the Sun as a reference for our analysis.

The abundance determination was performed by using the line-by-line differential method, employing the EW manually measured by fitting Gaussian profiles with the IRAF {\it splot} task and deblending when necessary.
Very special care was taken for the continuum placement during the measurements, always comparing and overplotting the spectral lines region for the sample, focusing on a consistent determination.

With the measured EW, we first determined the Fe I and Fe II abundances to differentially obtain the stellar parameters. For this we employed the 2014 version of the LTE code MOOG \citep{sne73} with the MARCS grid of 1D-LTE model atmospheres \citep{gus08}. It is important to highlight that the choice of a particular atmospheric model has a minor impact on the determination of stellar parameters and chemical abundances in a strictly differential analysis, as long the stars that are being studied are similar to the star of reference \citep[e.g.][]{tuc16}.

To make the analysis more efficient, we employed the Python q${\rm^2}$ code\footnote{https://github.com/astroChasqui/q2} \citep{ram14}, which operates in a semi-automatic mode, by calling MOOG routines to determine the elemental abundances and perform a line-by-line differential analysis using these results. This code also performs corrections of hyperfine structure (HFS) and the determination of uncertainties. 
In this work, we take into account the HFS for V, Mn, Co, Cu, Y, Ag, La and Pr using the line list from \cite{mel12}.
The errors are computed considering both observational (due to uncertainties in the measurements, represented by the standard error) and systematic uncertainties (from the stellar parameters and their inter-dependences), as described in in \cite{ram15}. Observational and systematic errors are added in quadrature.

Table \ref{param} shows the stellar parameters obtained for 16 Cyg A and B using the Sun (T$_{eff}$= 5777 K, log g = 4.44 dex, [Fe/H] = 0.0 dex) as reference.
Note that these results are practically the same within the errors as the ones found in \cite{tuc14}, which are T$_{eff}$ = 5830 $\pm$ 7 K, log g = 4.30 $\pm$ 0.02 and [Fe/H] = 0.101 $\pm$ 0.008 dex for
16 Cyg A, and T$_{eff}$ = 5751 $\pm$ 7 K, log g = 4.35 $\pm$ 0.02 and [Fe/H] = 0.054 $\pm$ 0.008 dex for 16 Cyg B.
The final difference in metallicity between the components of this binary system is remarkably similar to the one reported in \cite{tuc14}, that is $\Delta$ [Fe/H] = 0.047 $\pm$ 0.005 dex, while we find in this work $\Delta$[Fe/H] = 0.040 $\pm$ 0.006 dex. This confirms with a significance of $\sim 7\sigma$ that 16 Cyg A is indeed more metal rich when compared to 16 Cyg B, in agreement with \cite{ram11} and the earlier work by \cite{law01}, as well as the recent work by \cite{nis17}.

\subsection{Trigonometric surface gravity}

New parallaxes for the binary stars of 16 Cygni have been measured by the Gaia mission DR 2 \citep{gai18}. The new values are $47.2771 \pm 0.0327$ mas and $47.2754 \pm 0.0245$ mas for 16 Cyg A and B, respectively. Adopting the magnitudes from the General Catalogue of Photometric Data \citep[GCPD;][]{mer97} 
with V(A) = $5.959 \pm 0.009$  and V(B) = $6.228 \pm 0.019$, we determined the absolute magnitudes $M_{A} = 4.332 \pm 0.012$ and $M_{B} = 4.599 \pm 0.026$.
Using this information with the values of the T$_{eff}$, metallicity and mass, we estimate the trigonometric surface gravity for the pair of stars. For 16 Cyg A we found log g($A_{T}$) = 4.293 $\pm$ 0.005 dex and for 16 Cyg A log g($B_{T}$) = 4.364 $\pm$ 0.006. Notice that, while the surface gravity for 16 Cyg B has a good agreement with the one found through the ionization equilibrium of Fe lines (Table \ref{param}), for 16 Cyg A the trigonometric value is $\sim$ 0.02 dex lower, albeit they agree within 1.5 $\sigma$. Comparing with the results of \cite{ram11}, both the trigonometric and Fe-lines-based surface gravity are in agreement for the A component, while our surface gravities for B are somewhat higher in both cases.

\subsection{Age, mass and radius}

The age and mass of the binary stars were determined using customized Yonsei-Yale isochrones \citep{yi01}, as described in \cite{ram13,ram14}. 
This method provides good relative ages, due to the high precision achieved for the atmospheric
parameters. We estimate the ages and masses with probability distribution functions, through the comparison of atmospheric parameters position
of the star with the values predicted by the isochrones. 
Initially, the calculations were based on the [Fe/H], T$_{\rm eff}$ and log $g$, and later we replaced the gravity by the parallax values and magnitudes, to obtain the isochronal ages using the absolute magnitudes.
The results are shown in Table \ref{param}.
Our masses and radii shows a very good agreement when compared to the asteroseismology determinations of M$_{A} = 1.08 \pm 0.02 M_{\odot}$, M$_{B}= 1.04 \pm 0.02 M_{\odot}$,
R$_{A}= 1.229 \pm 0.008 R_{\odot}$ and R$_{B}= 1.116 \pm 0.006 R_{\odot}$, as reported by \cite{met15}.

\begin{table}
\centering
\caption{Stellar parameters for the 16 Cygni binary system using EW measured manually}
\label{param}
{\centering
\renewcommand{\footnoterule}{}
\begin{tabular}{lcc} 
\hline\hline
  {} & 16 Cyg A & 16 Cyg B\\

 \hline 
 T$_{eff}$ (K) & 5832$\pm$5 & 5763$\pm$5\\
 log g (dex) & 4.310$\pm$0.014& 4.360$\pm$0.014\\
 log g (dex)$_{trigonometric}$ & 4.293$\pm$0.005& 4.364$\pm$0.006\\
 $[$Fe/H$]$ (dex) & 0.103$\pm$0.004 & 0.063$\pm$0.004\\
 $v_t$ (km.s$^{-1}$) &1.11$\pm$ 0.01 & 1.03$\pm$ 0.01\\
 Luminosity(L$\odot$)$_{\log g}$ &1.46 $\pm 0.05$ & 1.19 $\pm 0.04$ \\
 Luminosity(L$\odot$)$_{parallax}$ & 1.55 $\pm 0.02$ & 1.23 $\pm 0.02$ \\
 Mass (M$\odot$)$_{\log g}$ & $1.06 \pm 0.02$& $1.01 \pm 0.01$\\
 Mass (M$\odot$)$_{parallax}$ & $1.06 \pm 0.01$& $1.01 \pm 0.01$\\ 
 Radius (R$\odot$)$_{\log g}$ & $1.19 \pm 0.02$ & $ 1.09 \pm 0.02$\\
 Radius (R$\odot$)$_{parallax}$ & $1.18 \pm 0.02$ & $ 1.12\pm 0.01$\\
 Age (Gyr)$_{\log g}$ &$6.0 \pm 0.3$ & $6.8 \pm 0.4$\\
 Age (Gyr)$_{parallax}$ &$6.4 \pm 0.2$ & $7.1 \pm 0.2$\\
 Age (Gyr)$_{[Y/Mg]}$ &$6.2 \pm 1.0$ & $6.3 \pm 1.0$\\
 Age (Gyr)$_{[Y/Al]}$ &$6.6 \pm 1.0$ & $6.8 \pm 1.0$\\
 
 \hline   
 \\
 \end{tabular}
 }
\end{table}

The inferred isochronal ages of A and B based on log $g$, are 6.0 $\pm$ 0.3 Gyr and 6.7 $\pm$ 0.4 Gyr, respectively. This shows that both components of the system have roughly the same age (within error bars). 
We have also estimated the ages of 16 Cyg A and B using the correlation of [Y/Mg] and [Y/Al] as a function of stellar age. The abundance clock [Y/Mg] for solar-type stars was first suggested by \cite{sil12} and the correlation between [Y/Mg] or [Y/Al] and stellar age was quantified for solar twins 
by \cite{nis15}, \cite{tuc16} and \cite{spi16} \footnote{Notice that the correlation between [Y/Mg] and age is only valid for solar-metallicity stars \citep{fel17}}.
The derived [Y/Mg] ages are A= 6.2 Gyr and B= 6.3 Gyr using the relation of \cite{tuc16}. 
Similar ages are found using the \cite{spi16} relations, A= 6.0 $\pm 1.0$ Gyr and B= 6.1 $\pm 1.0$ Gyr. 
These results are consistent with the values calculated using the isochronal method, while the [Y/Al] ages \citep{spi16} give A =6.6 $\pm 1.0$ Gyr and B= 6.8 $\pm 1.0$ Gyr.
The values of age, mass and radius are in agreement with
asteroseismic with values around 7 Gyrs \citep{san16, bel17}, with 16 Cyg A being slightly more massive and with bigger radius than its companion.

\subsection{Activity}

The chromospheric activity is an important constrain on stellar ages \citep{lor18}. In order to measure the activity differences between 16 Cyg A and B, we defined an instrumental activity index based on H$\alpha$ line which is a well-known chromospheric indicator of late-type stars \citep{pasquini91,lyra05,montes01}:
\begin{equation}\label{eq:haindex}
\mathcal{H} = \frac{F_{\rm H\alpha}}{(F_{\rm B}+F_{\rm V})},
\end{equation}

where $F_{\rm H\alpha}$ is the flux integrated around the H$\alpha$ line ($\Delta\lambda$ = 6562.78 $\pm$ 0.3 \AA). We chose this narrow spectral interval to minimize the effective temperature effects that might be present along the H$\alpha$ wings\footnote{Small residual photospheric effects are still expected to be affecting our index measurements, however, this residual feature should have negligible impact on our results since we are not interested in absolute activity scale determination for a wide range of effective temperatures.}. $F_{\rm B}$ and $F_{\rm V}$ are the fluxes integrated around 0.3 \AA\ continuum windows, centered at 6500.375 and 6625.550 \AA, respectively. In table \ref{table:haindex}, we show the estimated $\mathcal{H}$ for 16 Cyg AB and the Sun. The uncertainties were estimated by quadratic error propagation of equation \ref{eq:haindex}, assuming Poisson error distribution.  

\begin{table}
\caption{Activity indexes for 16 Cyg A, B, and the Sun. The last row is the mean activity level of 16 Cyg AB.}
 \begin{center}
\begin{tabular}{l | c}
  \hline
 Star & $\mathcal{H}$ \\ \hline  \hline
Sun & 0.1909 $\pm$ 0.0019 \\
16 Cyg A & 0.1871 $\pm$ 0.0021 \\
16 Cyg B & 0.1889 $\pm$ 0.0021 \\
 \hline
16 Cyg AB & 0.1880 $\pm$ 0.0010 \\
 \hline
 \end{tabular}
\end{center}
\label{table:haindex}
\end{table}

Accordingly to $\mathcal{H}$, none of 16 Cyg components show unexpected level of chromospheric activity for a typical 6-7 Gyr-old star \citep{mamajek08}. Furthermore, 16 Cyg A and B seem to be chromospherically quiet stars ($\mathcal{H}$ = 0.188 $\pm$ 0.001) and slightly more inactive than the Sun ($\mathcal{H}$ = 0.1909 $\pm$ 0.0019), indicating a chromospheric age older than 4-5 Gyr. This result is in line with Ca II H \& K multi-epoch observations of \citet{isaacson10} who found $\log(R^\prime_{\rm HK})$ $\approx$ -5.05 dex for this system, in good agreement with the mean activity level of $\log(R^\prime_{\rm HK})$ = -5.03 $\pm$ 0.1 dex derived for 49 solar-type stars from the 6-7 Gyr old open cluster NGC 188 \citep{lorenzo16b}. 

We inspected the chromospheric signature of other classical indicators along the spectral coverage of our observations such as Ca II H \& K \citep{mamajek08,lorenzo16b}, H$\beta$ \citep{montes01} and Ca II infrared triplet \citep{lorenzo16}. All of them show the same behavior found by H$\alpha$ lines, 16 Cyg A and B are chromospherically older than the Sun (age $>$ 4-5 Gyr) and the activity differences between the components are negligible. In summary, the different activity indicators reinforce the
age results from isochrones and seismology.

\section{Abundance analysis}

We present high-precision abundances for the light elements C, N, O, Na, Mg, Al, Si, S, K, Ca, Sc, Ti, V, Cr, Mn, Co, Ni, Cu and  Zn; and the heavy elements
Sr, Y, Zr, Ba, Ru, Rh, Pd, Ag, La, Ce, Nd, Sm, Eu, Gd, and Dy. The abundances of these elements were differentially determined using
initially the Sun as our standard star and then using 16 Cyg B as the reference to obtain the $\Delta [$X/H$]_{(A-B)}$.
The calculations were performed with the same method as described for the iron lines (see also \cite{tuc16}).

\begin{figure}
\centering
\includegraphics[width=1.0\columnwidth]{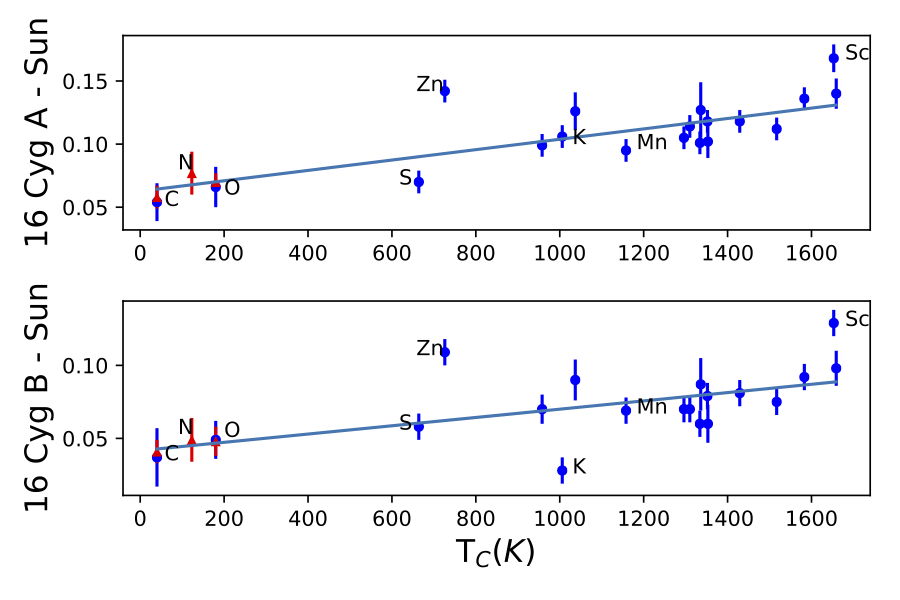}
\caption{Elemental abundances of 16 Cyg A (upper panel) and B (lower panel) based on our manually measured solar abundances as a function of condensation temperature for light elements (Z$\leq 30$). Solid lines
represent the linear fits with a slope of $3.99 \pm 0.58\times10^{-5} $ for the A component and $2.78 \pm 0.57\times10^{-5}$ for 16 Cyg B.
Red triangles correspond to the molecule-based abundances of C, N, and O.}
\label{cyg_sol}
\end{figure}

Taking into account only the elements with Z $\leq 30$, there is a clear chemical trend as a function of the condensation temperature (T$_{c}$) in the pattern of both stars relative to the Sun (Figure \ref{cyg_sol}), in agreement with \cite{ram11}, \cite{sch11} and \cite{tuc14}. In addition to results based on atomic lines, abundances for the volatiles elements C, N, and O were also determined using the molecules CH, NH and OH (red triangles in Figures \ref{cyg_sol} and \ref{cygab}). There is a very good agreement between the C and O abundances based on high excitation atomic lines and low excitation molecular lines, while for N we only present the abundance based on NH.
The excellent agreement between atomic and molecular-based differential abundances reinforces the reliability of our adopted atmospheric parameters.

\subsection{Abundance vs. condensation temperature trend}

A possible indication of rocky planet formation (or planet engulfment) can be found in the distribution of the differential elemental abundances as a function of condensation temperature. Refractory elements have high condensation temperature (T$_{C} \gtrsim $ 900 K), easily forming dust, being thus an important component of rocky bodies. Terrestrial planets (or the core of giant planets) may influence the surface abundance of its host
star in two ways: ${\it i)}$ the accretion of rocky material (planetary engulfment) depleted of hydrogen that enrich the stellar atmosphere in refractories
\citep[e.g.,][]{spi15, mel17, pet17}; ${\it ii)}$ imprisonment of refractory rich material into rocky objects (i.e, planetesimals, rocky planets, core of giant planets), that deplete the material accreted by the star during its formation \citep[e.g.,][]{mel09, ram11, tuc14}. In the case of planet engulfment, the thermohaline mixing should dilute the overabundance in a few milion year \citep{the12}, however, the thermohaline mixing could not be as effective and still leave some enhancement on the outer layers of the star, that can only be detected with a precision of $\sim 0.01$ dex.

An important point to highlight is that the signature of planet formation or planet engulfment is directly connected to the size of the convective zone during the event. If a solar-mass protostar would go through a fully convective phase that would last longer than the lifetime of the protoplanetary disk \citep[as it is conventionally accepted;][]{hay61}, any event of planetary formation that occur during such phase would be masked by a significant dilution with the stellar material enclosed into the convective zone, which would homogenize the chemical content throughout the star \citep[see Figure 2 of][]{spi15}.

In contrast to the classic steady accretion, there is the scenario of episodic material accretion onto the star (with observational evidence reported by \cite{liu16b}). Models that include episodic accretion can reach the stabilization of the convective zone earlier than 10 Myr with initial mass of $10 M_{Jup}$ and accretion rate bursts of $5 \times 10^{-4} M_{\odot}$yr$^{-1}$, reaching a final mass of 1 $M_{\odot}$ \citep{bar10}. Although is an extreme of their models, it is important to highlight that due to the effects of episodic accretion, the higher the mass of the accretion rate bursts for a given initial mass (or lower the initial mass for a given accretion rate) the greater the impact on the internal structure, reaching the necessary central temperature for the development of the radiative core ($\sim 2 - 3 \times 10^{6}$ K) earlier than what is predicted by the model of a non accreting star \citep[see Fig. 2 and Fig. 4 of][]{bar10}. This effect makes plausible the assumption that the formation of rocky bodies can chemically alter the surface abundance pattern of its parent star. 

Following this premise, \cite{mel09} suggested that the depletion of refractory elements in the Sun, when compared to a sample of 11 solar twins (without information regarding planets), is due to the formation of terrestrial planets in the solar system (see also further work by \cite{ram09, ram10}). However, in the literature, there are different suggestions for the Sun's abundance trend with condensation temperature. 

\cite{adi14} proposed that the trend with condensation temperature is an effect of the chemical evolution of the Galaxy or depends on the star's birthplace. 
Investigating the influence of age on solar twins, \cite{nis15} found a strong correlation of $\alpha$ and s-process elements abundances with stellar age, findings that were confirmed by \cite{tuc16} and \cite{spi16}.
According to \cite{one14}, if the star is formed in a dense stellar environment, the gas of the proto-stellar disk could have its dust cleansed before its birth by radiation of hot stars in the cluster, but recent theoretical estimates by \cite{gus18}, suggest that the mechanism is not significant.  
\cite{gai15} associate this effect to the gas-dust segregation in the protoplanetary disk. 
\cite{mal16} find differences in the T$_{c}$-slopes of refractory elements between stars with and without known planets, but this effect depends on the evolutionary stage, since it has been detected on main-sequence and subgiant stars, while no trend is found in their sample of giants. The authors also suggest that there is a correlation of both the mass and age, with T$_{c}$.

In this context, the investigation of abundance peculiarities in binary stars with and without planets is essential, because in a binary system there is no effect due to the chemical evolution of the Galaxy and other external factors, because it would equally affect both stars and thus be minimized in a differential analysis.
In this sense, the 16 Cygni system is a very interesting case, where both components are solar twins with the same age from asteroseismology \citep{san16}. On top of that, 16 Cyg B has a detected giant planet with a minimal mass of 1.5 Jupiter mass \citep{coc97}
while 16 Cyg A has no planet detected up to now, being thus a key target to study the effect of planets on the chemical composition of stars.

However, the abundance pattern of 16 Cyg A relative to B is still a controversy. A few authors suggest that there is no difference on the metallicity of the pair \citep{del00, sch11, tak11}, while most found abundance differences of about 0.05 dex \citep{nis17, adi16b, mis16, tuc14, ram11, law01, gon98}.

\subsection{16 Cygni}

A linear fit was performed with orthogonal distance regression (ODS) using the individual abundance errors for each element, excluding K due to its uncertain non-LTE effects. 
It was necessary to assume a minimum threshold for the abundances uncertainties because some species were returning very small error bars (0.001 dex), heavily impacting the abundance vs. condensation temperature slope, because some species does not have many lines. In order to address this issue, we adopt a minimum abundance error of 0.009 dex, which is the average error of all species analyzed.

We obtain the slopes $3.99 \pm 0.58\times10^{-5}$ dex K$^{-1}$ and $2.78 \pm 0.57\times10^{-5}$ dex K$^{-1}$ for 16 Cyg A $-$ Sun, and 16 Cyg B $-$ Sun, vs condensation temperature, respectively. 
In contrast to our past work \citep{tuc14}, we do not break the linear fit into two distinct curves for the volatiles and refractory elements, as a simple linear fit represents well the trend with T$_{c}$. We include nitrogen from NH, and for the abundances of C and O we assumed the average between the molecular and atomic abundances.

\begin{figure}
\centering
\includegraphics[scale=1.2,width=1.0\columnwidth]{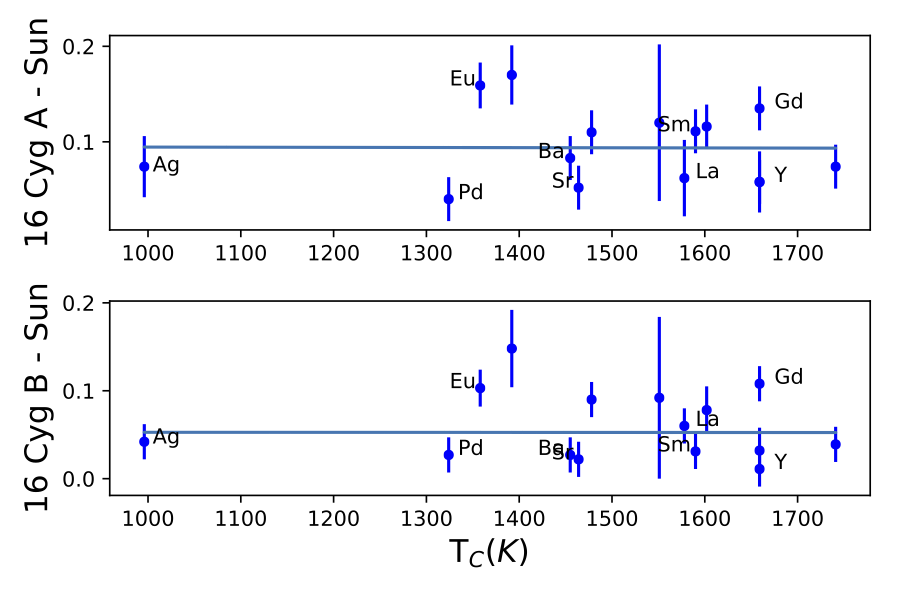}
\caption{As Figure \ref{cyg_sol} but for the heavy elements (Z $>$ 30). There is no clear trend with condensation temperature.}
\label{heavy_sol}
\end{figure}

In Figure \ref{heavy_sol} we plot the abundances of the heavy elements (Z $>$ 30). In this case, the abundances do not clearly follow the same trend as in the previous case, with slopes $-0.16 \pm 3.99 \times10^{-5}$ dex K$^{-1}$ and $-0.05 \pm 2.97 \times10^{-5}$ dex K$^{-1}$ for 16 Cyg A $-$ Sun and 16 Cyg B $-$ Sun, repectively, with a minimum uncertanty threshold of 0.02 dex in both cases.
However, due to the large errors in the [X/H] and the small range in T$_{c}$ it is not possible to claim if there is indeed a trend by considering only the heavy elements, but we stress that, within the uncertainties, the slope is not actually different from those of Z $\leq$ 30.
Although no T${c}$ trend is detected, there is a difference $\Delta$(A-B) = 0.043 $\pm$ 0.075 dex regarding the abundances of these heavy elements, that somewhat follow the difference of Fe between the stars of the pair, however, due to the high uncertainty we cannot conclude if this discrepancy is real.

\begin{figure*}
\centering
\includegraphics[width=1.9\columnwidth]{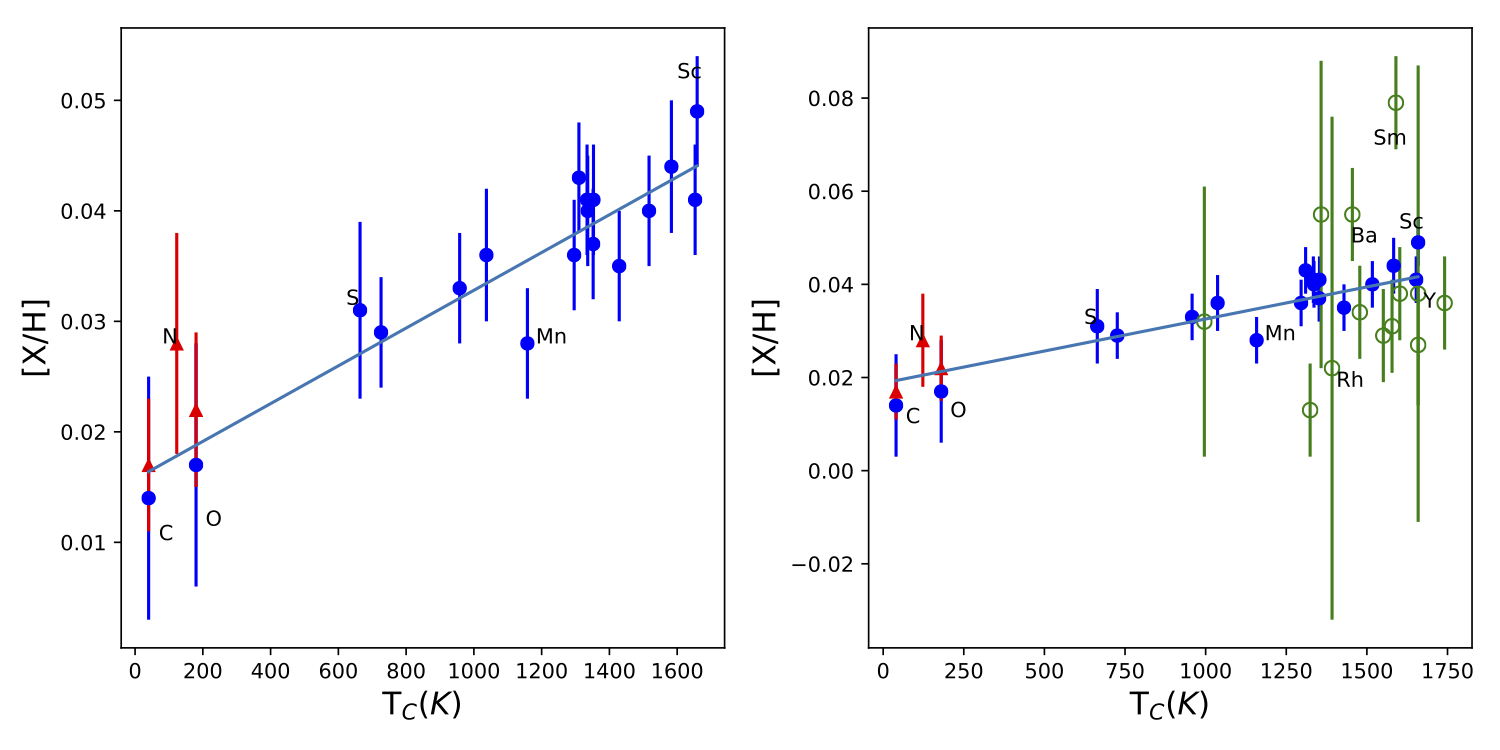}
\caption{Differential abundances (manually measured) of (A - B) as function of T$_{c}$ for elements with Z $\leq$ 30 (left panel) and adding also the neutron-capture elements (right panel). The red triangles correspond to the molecule-based abundances of C, N and O. The slope found is $1.56 \pm 0.24 \times  10^{-5}$ dex K$^{-1}$, based on the fit to the elements with Z $\leq$ 30.}
\label{cygab}
\end{figure*}

The abundances of 16 Cyg A relative to 16 Cyg B were also determined and are presented in Figure \ref{cygab}.
There is an evident trend between the (A-B) abundances and T$_{c}$.
The slope of the linear fit (without including the n-capture elements) is $1.56 \pm 0.24 \times 10^{-5}$ dex K$^{-1}$ (with a threshold of 0.005 dex). This result agrees with Tucci Maia et al. (2014; slope = $1.88 \pm 0.79 \times 10^{-5}$ dex K$^{-1}$) within error bars, showing once again the consistency and robustness of our analysis. If we include the heavy elements on the fit, we find a slope of $1.38 \pm 0.41 \times  10^{-5}$ (with a threshold of 0.010 dex).
Although the abundance of potassium presented in Table \ref{abun} has been corrected for non-LTE effects using the grid by \cite{tak02}, we did not use it for the linear
fit as the non-LTE grid is too sparse for a precise correction. Our slope is also in good agreement with the recent result by \cite{nis17}, $+ 0.98 \pm 0.35 \times 10^{-5}$ dex K$^{-1}$, , based on high-resolution high-S/N HARPS-N spectra.

\begin{table*}
\centering
\caption{Elemental abundances of 16 Cygni system relative to the Sun and to 16 Cyg B.}
\label{abun}
{\centering
\renewcommand{\footnoterule}{}
\begin{tabular}{lcccccccc} 
\hline\hline
  Z & Species & T$_{c}$& [X/H]$_{\text{16 Cyg A}}$ & Error & [X/H]$_{\text{16 Cyg B}}$ & Error & [X/H]$_{\text{A-B}}$ & Error\\
 \hline 
6  & C  &    40 & 0.049 & 0.013 & 0.033 & 0.004 & 0.012 & 0.009\\
6  & C* &    40 & 0.058 & 0.009 & 0.041 & 0.008 & 0.017 & 0.006\\ 
7  & N* &   123 & 0.077 & 0.017 & 0.049 & 0.015 & 0.028 & 0.010\\
8  & O  &   180 & 0.063 & 0.014 & 0.050 & 0.009 & 0.012 & 0.008\\
8  & O* &   180 & 0.070 & 0.007 & 0.048 & 0.010 & 0.022 & 0.007\\
11 & Na &   958 & 0.099 & 0.008 & 0.070 & 0.010 & 0.033 & 0.001\\
12 & Mg &  1336 & 0.127 & 0.022 & 0.087 & 0.018 & 0.040 & 0.005\\
13 & Al &  1653 & 0.168 & 0.011 & 0.129 & 0.009 & 0.041 & 0.004\\
14 & Si &  1310 & 0.114 & 0.008 & 0.070 & 0.007 & 0.043 & 0.003\\
16 & S  &   664 & 0.070 & 0.009 & 0.058 & 0.009 & 0.031 & 0.008\\
19 & K  &  1006 & 0.106 & 0.006 & 0.028 & 0.005 & 0.049 & 0.001\\
20 & Ca &  1517 & 0.112 & 0.002 & 0.075 & 0.002 & 0.040 & 0.003\\
21 & Sc &  1659 & 0.140 & 0.012 & 0.098 & 0.012 & 0.049 & 0.004\\
22 & Ti &  1583 & 0.136 & 0.003 & 0.092 & 0.002 & 0.044 & 0.006\\
23 & V  &  1429 & 0.118 & 0.006 & 0.081 & 0.003 & 0.035 & 0.004\\
24 & Cr &  1296 & 0.105 & 0.004 & 0.070 & 0.004 & 0.036 & 0.002\\
25 & Mn &  1158 & 0.095 & 0.005 & 0.069 & 0.006 & 0.028 & 0.003\\
26 & Fe &  1334 & 0.101 & 0.004 & 0.060 & 0.004 & 0.041 & 0.004\\
27 & Co &  1352 & 0.118 & 0.007 & 0.079 & 0.009 & 0.037 & 0.003\\
28 & Ni &  1353 & 0.102 & 0.013 & 0.060 & 0.013 & 0.041 & 0.004\\
29 & Cu &  1037 & 0.126 & 0.015 & 0.090 & 0.014 & 0.036 & 0.006\\
30 & Zn &   726 & 0.142 & 0.006 & 0.109 & 0.005 & 0.029 & 0.004\\
38 & Sr &  1464 & 0.052 & 0.003 & 0.022 & 0.002 & 0.030 & 0.004\\ 
39 & Y  &  1659 & 0.058 & 0.008 & 0.011 & 0.010 & 0.041 & 0.005\\ 
40 & Zr &  1741 & 0.074 & 0.017 & 0.039 & 0.018 & 0.036 & 0.003\\ 
44 & Ru &  1551 & 0.120 & 0.082 & 0.092 & 0.092 & 0.029 & 0.010\\ 
45 & Rh &  1392 & 0.170 & 0.031 & 0.148 & 0.044 & 0.022 & 0.054\\ 
46 & Pd &  1324 & 0.040 & 0.007 & 0.027 & 0.011 & 0.013 & 0.006\\ 
47 & Ag &   996 & 0.074 & 0.032 & 0.042 & 0.003 & 0.032 & 0.029\\ 
56 & Ba &  1455 & 0.083 & 0.005 & 0.027 & 0.003 & 0.055 & 0.001\\ 
57 & La &  1578 & 0.062 & 0.040 & 0.060 & 0.010 & 0.031 & 0.010\\ 
58 & Ce &  1478 & 0.110 & 0.020 & 0.090 & 0.011 & 0.034 & 0.007\\ 
60 & Nd &  1602 & 0.116 & 0.021 & 0.078 & 0.027 & 0.038 & 0.008\\ 
62 & Sm &  1590 & 0.111 & 0.009 & 0.031 & 0.011 & 0.079 & 0.007\\ 
63 & Eu &  1358 & 0.159 & 0.024 & 0.103 & 0.021 & 0.055 & 0.033\\ 
64 & Gd &  1659 & 0.135 & 0.016 & 0.108 & 0.007 & 0.027 & 0.013\\ 
66 & Dy &  1659 & 0.058 & 0.032 & 0.032 & 0.026 & 0.038 & 0.049\\ 
 
 \hline                                 

 \hline
 
 \end{tabular}
 }
\\
$^{\rm *}$ molecular abundance
\end{table*}

\subsection{Automated codes}

We conducted tests utilizing iSpec version 2016 \citep{bla14} and ARES v2 \citep{sou15} to automatically measure the EWs of 16 Cyg A, B and the Sun, in order to differentially  determine the stellar parameters. The aim of these tests is to evaluate if these codes, when applied to high-resolution data and following our methodology (same as in Section 2.2), could return a similar result to what we find by "hand". Our motivations for this is to assess if our procedure could be automatized and applied to a bigger sample of stars and still retrieve stellar parameters with the same precision as ours, and to also find out if the chemical composition differences that we found between the 16 Cygni components is consistent by applying different methods of EW meassurement. 

As discussed earlier, the differential method minimize most of the error sources while, for a solar twin sample, the uncertainty is almost entirely related to the EW. One big concern in a differential analysis is the continuum normalization to achieve a consistent continuum placement for all stars being analyzed.
In this test, the spectra is the same as our manual analysis, which was previously normalized, and the EW measured following the same line list as ours.
In this way, any discrepancy in the values would be due to how each code interprets and places the continuum, and how the fit is performed.

\begin{table*}
\centering
\caption{Atmospheric parameters for 16 Cygni determined with automated EW measurements for R= 160 000 and 81000 spectra.}
\label{param_au}
{\centering
\renewcommand{\footnoterule}{}
\begin{tabular}{lllllll} 
\hline\hline
{} & {R= 160 000}& {} & {R= 81 000} & {}\\
\hline
{} & A & B & A & B\\
\hline
{iSpec} & {} & {} & {} & {}\\
T$_{eff}$ (K) & 5834 $\pm$ 5 & 5749 $\pm$ 4 &  5826$\pm$15 & 5753$\pm$14\\
log $g$ (dex) & 4.330 $\pm$ 0.013& 4.360 $\pm$ 0.011 & 4.320$\pm$0.042& 4.350$\pm$0.044\\
$[$Fe/H$]$ (dex) & 0.103 $\pm$ 0.005 & 0.052 $\pm$ 0.003 & 0.098$\pm$0.013 & 0.057$\pm$0.013\\
v$t$ (km.s$^{-1}$) &  1.09$\pm$ 0.01 & 1.02$\pm$ 0.01 &  1.11$\pm$ 0.03 & 1.03$\pm$ 0.03\\
\hline
{ARES} & {} & {} & {} & {}\\
T$_{eff}$ & 5840$\pm$16 & 5753$\pm$15 & 5813$\pm$26 & 5760$\pm$19\\
log $g$ (dex) & 4.330$\pm$0.040& 4.320$\pm$0.038 & 4.290$\pm$0.063& 4.390$\pm$0.055\\
$[$Fe/H$]$ (dex) & 0.114$\pm$0.013 & 0.048$\pm$0.012 & 0.091$\pm$0.022 & 0.061$\pm$0.017\\
v$t$ (km.s$^{-1}$) & 1.07$\pm$ 0.03 & 1.02$\pm$ 0.03 & 1.06$\pm$ 0.05 & 0.97$\pm$ 0.04\\
\hline

\hline
 
 \end{tabular}
 }
\end{table*}

In addition to that, we use another set of spectra with lower resolving power (R $\sim$ 81 000, from \cite{tuc14}), in order to evaluate if by using the same tools but providing different resolution spectra, the results could be somewhat discrepant.
In Table \ref{param_au} we present the determined stellar parameters obtained using the EW measurements obtained with the automated codes. Comparing the results, we find that all the codes return stellar parameters in good agreement to ours, based on the "manual" measurements.
In overall, as we go to lower resolution, the uncertainty gets higher, as expected, because it can lead to more blends around the lines measured and thus a more contaminated value for the EW, a phenomenon that also happens with manual measurements as well, as we can see from the stellar parameters from our previous work \citep{tuc14}.

\begin{table*}
\centering
\caption{Slopes of abundances versus condensation temperature, for the elements with Z $\leq$ 30 for the EWs measurements from iSpec and ARES for the R $\sim$ 160 000 and 81 000 spectra.}
\label{slopab_sig}
{\centering
\renewcommand{\footnoterule}{}
\begin{tabular}{clcclcc} 
\hline\hline
R & iSpec (dex K$^{-1}$) & Min. error (dex) & Significance & ARES (dex K$^{-1}$) & Min. error (dex) & Significance\\
160 000 & {$1.11 \pm 0.31\times10^{-5}$} & 0.005 & 3.58 &{$0.70 \pm 0.64\times10^{-5}$} & 0.010 & 1.09\\
81 000 & {$1.56 \pm 0.44\times10^{-5}$} & 0.008 & 3.55& {$1.24 \pm 1.04\times10^{-5}$} & 0.019 & 1.19\\
 
 \hline
 
 \end{tabular}
 }
\end{table*}

We also used iSpec and ARES to determine the elemental abundance of our sample for the elements with Z $\leq$ 30, with the same method and spectra described in the previous section. The trends with condensation temperature are presented in Table \ref{slopab_sig} with its respective abundance thresholds. In both resolution sets iSpec and ARES returns a T$_{c}$ slope that is in agreement with our value within error bars. The codes confirm not only that 16 Cyg A is more metal rich than B, but also the existence of T$_c$ trend even though we use spectra with almost half the resolving power (but still high-resolution spectra) as the other. However, iSpec shows a higher significance on its values.

\subsection{Li and Be}

Lithium and beryllium abundances were determined by performing spectral synthesis calculations, using a method similar to the outlined in \cite{tuc15}. For lithium, we used the Li-7 doublet at 670.7 nm and, for beryllium, we used the doublet resonance lines of Be II at 313.0420 nm and 313.1065 nm. The line list for the Li synthesis is from \cite{mel12}, while for Be we used a modified list of \cite{ash05}, as described in \cite{tuc15}.

For the spectral synthesis, we used the synth driver of the 2014 version of the 1D LTE code MOOG \citep{sne73}.
We adopted A(Be) = 1.38 dex as the standard solar Be abundance from \cite{asp09}.
The model atmospheres were interpolated from the MARCS grid \citep{gus08} using the stellar parameters previously obtained. The abundances of Li were corrected for non-LTE effects using the online grids of the INSPECT project\footnote{http://inspect-stars.com/}.
Beryllium lines are insensitive to non-LTE effects in the solar type stars, according to \cite{asp05}.

To determine the macroturbulence line broadening, we first analyzed the line profiles of the Fe I 602.7050 nm, 609.3644 nm, 615.1618 nm, 616.5360 nm, 670.5102 nm and Ni I 676.7772 lines in the Sun; the synthesis also included a rotational broadening of v sin i = 1.9 km.s$^{-1}$ \citep{bru84} and the instrumental broadening. The macroturbulent velocity found for the Sun is V$_{macro}$ = 3.6 km s$^{-1}$. For 16 Cygni, we estimate the macroturbulence following the relation of \cite{leo16}, which takes into account the dependence with effective temperature and log g.

With the macroturbulence fixed, v $\sin i$ was estimated for 16 Cyg A and B by fitting the profiles of the six lines mentioned above, also including the instrumental broadening.
Table \ref{libe} shows the abundances of Li and Be with their estimated macroturbulence and v $\sin i$.

\begin{table}
\centering
\caption{Abundances of Li and Be for the binary 16 Cyg using spectral synthesis}
\label{libe}
{\centering
\renewcommand{\footnoterule}{}
\begin{tabular}{lcc} 
\hline\hline
  {} & 16 Cyg A & 16 Cyg B\\
 \hline 
 Li (dex) & 1.31$\pm$0.03 & 0.61$\pm$0.03\\
 Be (dex)& 1.50$\pm$0.03& 1.43$\pm$0.03\\
 V$_{macro}$ (km s$^{-1}$) & 3.97 $\pm$ 0.25 & 3.66 $\pm$ 0.25\\
 v $\sin i$ (km s$^{-1}$) & 1.37 $\pm$ 0.04 & 1.22 $\pm$ 0.06\\

 \hline                                 
 \end{tabular}
 }
\end{table}

\begin{figure}
\centering
\includegraphics[width=1.0\columnwidth]{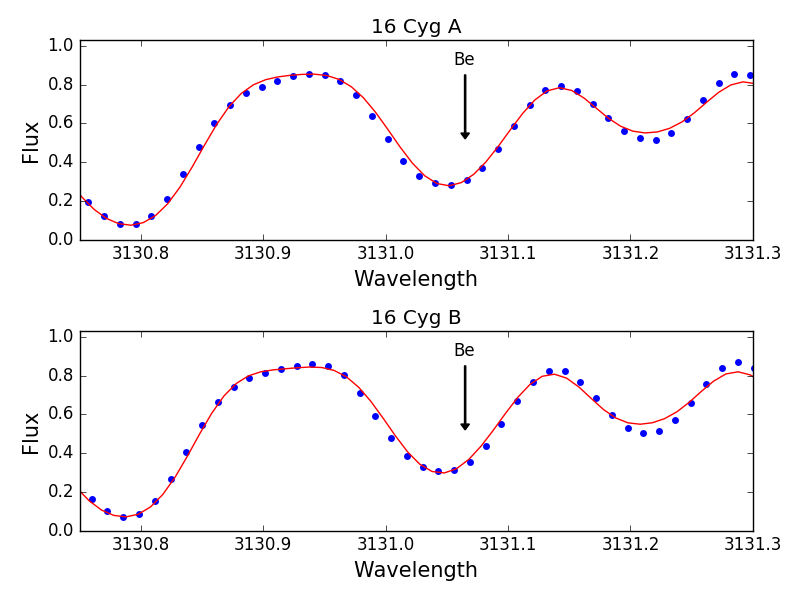}
\caption{ Comparison between the observed (blue dots) and synthetic (red solid line) spectra of 16 Cyg A (top) and 16 Cyg B (bottom).}
\label{beab}
\end{figure}

In Figure \ref{beab} we show the synthetic spectra of 16 Cyg A and B plotted against the observed spectra.

Comparing our results with previous works in the literature, we found that Li and Be on the binary system 16 Cygni is hardly a consensus, but for
lithium there is a qualitative agreement in the A component being more abundant in lithium than the B component.
What is important to highlight here is that we found a higher Be abundance on the A component when compared to B, in contrast to the results of \cite{del00} and \cite{gar98}, while \cite{tak11} does not find any significant Be variation between components, maybe because of the different parameters (including broadening parameters) and different spectra (resolving power, S/N, normalization). 

\begin{table*}
\centering
\caption{Comparison of Li and Be abundances.}
\label{libec}
{\centering
\renewcommand{\footnoterule}{}
\begin{tabular}{ccccc} 
\hline\hline
Star & HD & Li(dex) & Be(dex) & references\\

\hline

16 Cyg A & 186408 & 1.31 & 1.50 & ours\\
{-} & {-} & 1.37 & 1.34 & \cite{tak11}\\
{-} & {-} & 1.27$^{a}$ & 0.99$^{b}$ &$^{a}$\cite{kin97}; $^{b}$\cite{del00}\\
{-} & {-} & 1.24$^{c}$ & 1.10$^{d}$ & $^{c}$\cite{gon98}; $^{d}$\cite{gar98}\\
{-} & {-} & 1.34$^{d}$ & - &$^{d}$\cite{ram11}\\
16 Cyg B & 186427 & 0.61 & 1.43 & ours\\
{-} & {-} & $< 0.60$ & 1.37 & \cite{tak11}\\
{-} & {-} & $\leq 0.60^{a}$ & 1.06$^{b}$ & $^{a}$\cite{kin97}; $^{b}$\cite{del00}\\
{-} & {-} & $< 0.50^{c}$ & 1.30$^{d}$ & $^{c}$\cite{gon98}; $^{d}$\cite{gar98}\\
{-} & {-} & 0.73$^{d}$ & - &$^{d}$\cite{ram11}\\
 
 \hline                                 
 \end{tabular}
 }
\end{table*}

\section{Discussion}

Lithium and beryllium are elements that are destroyed at different temperatures ($2.5\times10^{6}$ K and $3.5\times10^{6}$ K, respectively) and therefore at different depths in the stellar interiors. According to standard stellar evolution models, these temperatures are only achieved below the base of the convective zone. However, the solar photospheric Li abundance is approximately 150 times lower than the meteoritic value, indicating that extra mixing processes are acting in solar-type stars and need to be taken into account.

In solar-type stars, it is known that Li has a strong correlation with age and surface rotation \citep{bec17, mar16, bau10, nas09}, which suggests internal depletion of lithium.
However, it is a more challenging task to do the same analysis for Be due to difficulties related to its detection utilizing instruments from the ground, 
with only two accessible lines of Be II being near the atmospheric cutoff in the UV, at 313 nm, in a heavily populated region of the spectrum.  
In \cite{tuc15}, we determined the abundance of Be in a sample of 8 solar twins through a ``differential'' spectral synthesis, where the line list was calibrated to match the observed solar spectrum, which was observed with the same setup as the other stars. We found that the Be content of solar twins is barely depleted, if at all, during the Main Sequence ($\sim 0.04$ dex in a time span of 8 Gyrs).
Thus, in a probable scenario of a planet being engulfed by its host star, if this event happens after the stabilization of its convective zone, one could expect an enhancement of Li and Be, in a similar way as for refractory elements. 

If we analyze our result with this hypothesis in mind, the overabundance of lithium and beryllium on 16 Cyg A relative to B (in addition to the enhancement of refractory elements) could indicate an accretion of mass.
In fact, \cite{san02} suggests that the Li abundance can be used as a signal of pollution enrichment of the outer layers of solar-type stars, if stellar ages are well known.  

The majority of previous studies agree that both components of the binary system have the same stellar ages. However, as seen in Table \ref{libec}, we found that 16 Cyg A is 0.70 dex more rich in Li than 16 Cyg B, in accord with the results of \cite{tak11, kin97, gon98}, and  \cite{ram11}. Furthermore, on the lithium-age trend of \cite{mar16} and \cite{mon13}, 16 Cyg B shows a normal Li abundance for a solar twin of its age, while 16 Cyg A has a Li abundance above the curve, thus, when compared to a sample of solar twins, the A component shows an anomalous abundance of lithium. On top of that, 16 Cyg A also seems to have a higher $v \sin i$ velocity (Table \ref{libe}), which may indicate momentum transferred by mass accretion. 

\cite{gon98} also suggests that the odd lithium abundance of 16 Cyg A may be due to planet accretion of a 1-2 Jupiter mass planet. This would increase the abundance not only of Li but of Fe as well. This is reinforced by \cite{maz97} who propose that the separation between the two stars \citep[semi-major axis of 755 AU;][]{pla13} is sufficiently small to permit planet-planet interactions to cause orbit instabilities on the binary pair. Furthermore, the high eccentricity \citep[0.689;][]{wit07} of 16 Cyg Bb could also be evidence of the interaction between the stars.
Similar results were found by \cite{law01}, who find a difference of 0.025 $\pm$ 0.009 dex in [Fe/H] between the 16 Cygni pair (the A component being more metal rich), suggesting a self-pollution scenario.
\cite{gra01} also investigated abundance differences in six main-sequence binaries with separations on the order of hundreds AU (enough to permit orbit instabilities on possible exoplanets) with components with almost the same mass, using the differential abundance technique (errors of the order of 0.01 dex). Four of these systems did not show any chemical differences between components, while the two remaining binary systems (HD 219542 and HD 200466) display a clear metallicity difference, being the primary stars more rich in iron (and in the most analyzed elements) than the secondary. The authors also support the idea that the difference in chemical composition of those binary stars is due to infall of rocky material.

By taking into account the hypothesis of planet accretion pollution, one could expect that the Be abundance would also be enriched on the outer layers of the star, in a similar way as Li. As discussed earlier, according to \cite{tuc15} beryllium is not depleted in a very effective way (if it is at all) on solar twins stars during the Main Sequence, making it also a good proxy for planet accretion after the stabilization of the convective zone. 
Comparing the pair of stars, we found that 16 Cyg A has 0.07 $\pm$ 0.03 dex more beryllium than 16 Cyg B, in line with the planet engulfment hypothesis.

Following the procedure outlined in \cite{gal16}, we estimate that, if we add 2.5 - 3.0 earth masses of Earth-like material into the convective zone of 16 Cyg B, we would alter the content of Be in about 0.07 dex, thus canceling the abundance difference between the stars. This estimate is close to the one derived by \cite{tuc14}, who derived that the addition of 1.5 earth mass of a material with a mixture of the composition of the Earth and CM chondrites is necessary to reproduce the refractory elements abundances as function of the condensation temperature pattern on 16 Cyg B. However, \cite{tuc14} assumed that this abundance pattern is a spectral signature of the 16 Cyg Bb rocky core formation and, now considering also the abundances of Li and Be, it may be a signature of planet accretion rather than planet formation.

In contrast, \cite{the12} discuss that engulfment of rocky planets can induce instabilities on a stellar surface, by the dilution of a metal-rich material in young Main Sequence stars, which creates an unstable $\mu$-gradient at the bottom of the convective zone, activating fingering (thermohaline) convection. 
This would be responsible for the depletion of the abundances that enriched the convective zone, thus quenching any signature of accretion. However, the authors also discuss that the mixing process would not completely erase the enhanced abundances, meaning that the engulfment event would still be detected in the high precision abundances domain. 
In this scenario, \cite{dea15} argues that, during the early periods on the Main Sequence, 16 Cyg B was able to accrete rocky material from its planetary disk, 
whereas no accretion may have developed around 16 Cyg A due to the presence of a red dwarf (16 Cyg C) orbiting at 73 AU around the A component \citep{tur01,pat02}.
The models of \cite{dea15}, that take into account the mixing by fingering convection, could reproduce the observed difference of lithium on the binary pair by adding 0.66 Earth mass to 16 Cyg B. However, in those same models, Be does not show any depletion with the addition of 0.66 Earth mass, with the destruction starting to be more effective with the accretion of higher masses. Notice that among the two dozen chemical elements showing abundance differences between 16 Cyg A and B, the model of \cite{dea15} can only explain the difference in lithium.

We find this scenario very unlikely because the lithium content of 16 Cyg B seems to be normal for its age when compared to other solar twins, while 16 Cyg A, on the other hand, is the one that displays an enhanced Li content for a solar twin with $\sim$ 7 Gyr \citep{mar16,mon13}. 

Another explanation for the discrepancy in Li abundances could be different initial rotation rates \citep{coc97}.
However, notice that although young solar-type stars of a given mass may have different rotation rates \citep{lor19}, they all seem to converge to the same rotation period at an age of about 0.2 Gyr \citep{bar03}, which is much earlier than the age of 16 Cyg.
Furthermore, the companion 16 Cyg C at 73 AU may be too far as to have any significant impact on 16 Cyg A.

\section{Conclusions}

We present a detailed study of elemental abundances on the 16 Cygni solar twin binary system using higher quality data (R = 160 000, S/N = 1000 at 670nm). We confirm the difference of 0.04 dex in [Fe/H] between 16 Cyg A and B. 
We also confirm the positive trend of differential abundances (A - B) as a function of condensation temperature, in very good agreement with our previous work \citep{tuc14}, which was obtained with spectra of lower resolving power and signal to noise ratio on a different instrument. There is also good agreement with the slope obtained independently by \cite{nis17}, and also using a different spectrograph (HARPS-N). 
We also find the same result by employing the ARES and iSpec codes to measure the EWs.
This shows that our differential analysis method is consistent and a powerful tool to unveil physical characteristics that can only be seen with high precision abundance determinations, and that the T$_c$ trend is a physical phenomenon, being thus unlikely to be related to some instrumental effect, as we show that high-quality spectra obtained with different spectrographs (Espadons at CFHT, HDS at Subaru, HARPS-N at the Telescopio Nationale Galileo) give essentially the same results (within error bars). 

We also determine the abundance of Li and Be through a ``differential'' spectral synthesis analysis, using the solar spectrum (obtained with the same instrumental configuration) to calibrate the line list that was used to perform the calculations. We found that 16 Cyg A exhibits an overabundance of not only Li (as reported by previous studies) but of Be as well, relative to 16 Cyg B. 
This discrepancy is compatible with a 2.5 - 3.0 Earth masses of earth like material if we assume a convective zone similar to Sun for both stars. 
Interestingly, the amount of rocky material needed to explain the Li and Be overabundances is also compatible
with the trend of the (A-B) abundances vs. condensation temperature, reinforcing thus the hypothesis of planet engulfment,
although a similar opposite trend in (B-A) could be attributed to the effect of the rocky core in 16 Cyg B \citep{tuc14}.
However, the overabundant Li content in 16 Cyg A, above what is expected for its age, suggest that the signature that we are observed is due to a planet engulfment event.

\begin{acknowledgements}
MTM acknowledges support by financial support of Joint committee ESO Chile and CNPq (312956/2016-9). 
JM, LS and DLO thanks FAPESP (2014/15706-9, 2016/20667-8, and 2018/04055-8) and CNPq (Bolsa de produtividade). 
PJ acknowledges FONDECYT Iniciación 11170174 grant for financial support.
\end{acknowledgements}



\begin{thebibliography}{}

\bibitem[Adibekyan et al.(2014)]{adi14} Adibekyan, V.~Z., Gonz{\'a}lez Hern{\'a}ndez, J.~I., Delgado Mena, E., et al.\ 2014, \aap, 564, L15 
\bibitem[Adibekyan et al.(2016)]{adi16b} Adibekyan, V., Delgado-Mena, E., Figueira, P., et al.\ 2016, \aap, 591, A34 
\bibitem[Ashwell et al.(2005)]{ash05} Ashwell, J.~F., Jeffries, R.~D., Smalley, B., et al.\ 2005, \mnras, 363, L81
\bibitem[Asplund(2005)]{asp05} Asplund, M.\ 2005, \araa, 43, 481
\bibitem[Asplund et al.(2009)]{asp09} Asplund, M., Grevesse, N., Sauval, A.~J., \& Scott, P.\ 2009, \araa, 47, 481
\bibitem[Baraffe \& Chabrier(2010)]{bar10} Baraffe, I., \& Chabrier, G.\ 2010, \aap, 521, A44
\bibitem[Barnes(2003)]{bar03} Barnes, S.~A.\ 2003, \apj, 586, 464 
\bibitem[Baumann et al.(2010)]{bau10} Baumann, P., Ram{\'{\i}}rez, I., Mel{\'e}ndez, J., Asplund, M., \& Lind, K.\ 2010, \aap, 519, A87
\bibitem[Beck et al.(2017)]{bec17} Beck, P.~G., do Nascimento, J.-D., Jr., Duarte, T., et al.\ 2017, arXiv:1702.01152
\bibitem[Bellinger et al.(2017)]{bel17} Bellinger, E.~P., Basu, S., Hekker, S., \& Ball, W.~H.\ 2017, \apj, 851, 80 
\bibitem[Biazzo et al.(2015)]{bia15} Biazzo, K., Gratton, R., Desidera, S., et al.\ 2015, \aap, 583, A135
\bibitem[Blanco-Cuaresma et al.(2014)]{bla14} Blanco-Cuaresma, S., Soubiran, C., Heiter, U., \& Jofr{\'e}, P.\ 2014, \aap, 569, A111 
\bibitem[Bruning(1984)]{bru84} Bruning, D.~H.\ 1984, \apj, 281, 830
\bibitem[Carlos et al.(2016)]{mar16} Carlos, M., Nissen, P.~E., \& Mel{\'e}ndez, J.\ 2016, \aap, 587, A100 
\bibitem[Chambers(2010)]{cha10} Chambers, J.~E.\ 2010, \apj, 724, 92
\bibitem[Cochram \& Hatzes(1997)]{coc97} Cochran, W.; Hatzes, A.; Butler, P.; Marcy, G.\ 2009, \apjl, 483, 457 
\bibitem[Deal et al.(2015)]{dea15} Deal, M., Richard, O., \& Vauclair, S.\ 2015, \aap, 584, A105
\bibitem[Deliyannis et al.(2000)]{del00} Deliyannis, C.~P., Cunha, K., King, J.~R., \& Boesgaard, A.~M.\ 2000, \aj, 119, 2437 
\bibitem[Do Nascimento et al.(2009)]{nas09} Do Nascimento, J.~D., Jr., Castro, M., Mel{\'e}ndez, J., et al.\ 2009, \aap, 501, 687 
\bibitem[Feltzing et al.(2017)]{fel17} Feltzing, S., Howes, L.~M., McMillan, P.~J., \& Stonkut{\.e}, E.\ 2017, \mnras, 465, L109 
\bibitem[Gaia Collaboration et al.(2018)]{gai18} Gaia Collaboration, Brown, A.~G.~A., Vallenari, A., et al.\ 2018, \aap, 616, A1.
\bibitem[Gaidos(2015)]{gai15} Gaidos, E.\ 2015, \apj, 804, 40
\bibitem[Galarza et al.(2016)]{gal16} Galarza, J.~Y., Mel{\'e}ndez, J., \& Cohen, J.~G.\ 2016, \aap, 589, A65 
\bibitem[Garcia Lopez \& Perez de Taoro(1998)]{gar98} Garcia Lopez, R.~J., \& Perez de Taoro, M.~R.\ 1998, \aap, 334, 599
\bibitem[Gonzalez(1998)]{gon98} Gonzalez, G.\ 1998, Brown Dwarfs and Extrasolar Planets, 134, 431
\bibitem[Gratton et al.(2001)]{gra01} Gratton, R.~G., Bonanno, G., Claudi, R.~U., et al.\ 2001, \aap, 377, 123 
\bibitem[Gustafsson et al.(2008)]{gus08} Gustafsson, B., Edvardsson, B., Eriksson, K., et al.\ 2008, \aap, 486, 951
\bibitem[Gustafsson(2018)]{gus18} Gustafsson, B.\ 2018, arXiv:1805.00547 
\bibitem[Hayashi(1961)]{hay61} Hayashi, C.\ 1961, \pasj, 13,
\bibitem[Isaacson \& Fischer(2010)]{isaacson10} Isaacson, H., \& Fischer, D.\ 2010, \apj, 725, 875 
\bibitem[King et al.(1997)]{kin97} King, J.~R., Deliyannis, C.~P., Hiltgen, D.~D., et al.\ 1997, \aj, 113, 1871 
\bibitem[Laws \& Gonzalez(2001)]{law01} Laws, C., \& Gonzalez, G.\ 2001, \apj, 553, 405 
\bibitem[Liu et al.(2014)]{liu14} Liu, F., Asplund, M., Ram{\'{\i}}rez, I., Yong, D., \& Mel{\'e}ndez, J.\ 2014, \mnras, 442, L51
\bibitem[Liu et al.(2016)]{liu16} Liu, F., Yong, D., Asplund, M., et al.\ 2016, \mnras, 456, 2636 
\bibitem[Liu et al.(2016b)]{liu16b} Liu, H.~B., Takami, M., Kudo, T., et al.\ 2016, Science Advances, 2, e1500875
\bibitem[Liu et al.(2018)]{liu18} Liu, F., Yong, D., Asplund, M., et al.\ 2018, \aap, 614, A138
\bibitem[Lorenzo-Oliveira et al.(2016)]{lorenzo16} Lorenzo-Oliveira, D., Porto de Mello, G.~F., Dutra-Ferreira, L., \& Ribas, I.\ 2016, \aap, 595, A11 
\bibitem[Lorenzo-Oliveira et al.(2016b)]{lorenzo16b} Lorenzo-Oliveira, D., Porto de Mello, G.~F., \& Schiavon, R.~P.\ 2016, \aap, 594, L3 
\bibitem[Lorenzo-Oliveira et al.(2018)]{lor18} Lorenzo-Oliveira, D., Freitas, F.~C., Mel{\'e}ndez, J., et al.\ 2018, \aap, 619, A73
\bibitem[Lorenzo-Oliveira et al.(2019)]{lor19} Lorenzo-Oliveira, D., Mel{\'e}ndez, J., Yana Galarza, J., et al.\ 2019, \mnras, 485, L68
\bibitem[Lyra \& Porto de Mello(2005)]{lyra05} Lyra, W., \& Porto de Mello, G.~F.\ 2005, \aap, 431, 329 
\bibitem[Mack et al.(2014)]{mac14} Mack, C.~E., III, Schuler, S.~C., Stassun, K.~G., \& Norris, J.\ 2014, \apj, 787, 98 
\bibitem[Mack et al.(2016)]{mac16} Mack, C.~E., III, Stassun, K.~G., Schuler, S.~C., Hebb, L., \& Pepper, J.~A.\ 2016, \apj, 818, 54
\bibitem[Maldonado \& Villaver(2016)]{mal16} Maldonado, J., \& Villaver, E.\ 2016, \aap, 588, A98
\bibitem[Mamajek \& Hillenbrand(2008)]{mamajek08} Mamajek, E.~E., \& Hillenbrand, L.~A.\ 2008, \apj, 687, 1264-1293 
\bibitem[Mazeh et al.(1997)]{maz97} Mazeh, T., Krymolowski, Y., \& Rosenfeld, G.\ 1997, \apjl, 477, L103 
\bibitem[Mel{\'e}ndez et al.(2009)]{mel09} Mel{\'e}ndez, J., Asplund, M., Gustafsson, B., \& Yong, D.\ 2009, \apjl, 704, L66 
\bibitem[Mel{\'e}ndez et al.(2012)]{mel12} Mel{\'e}ndez, J., Bergemann, M., Cohen, J.~G., et al.\ 2012, \aap, 543, A29
\bibitem[Mel{\'e}ndez et al.(2017)]{mel17} Mel{\'e}ndez, J., Bedell, M., Bean, J.~L., et al.\ 2017, \aap, 597, A34 
\bibitem[Mermilliod et al.(1997)]{mer97} Mermilliod, J.-C., Mermilliod, M., \& Hauck, B.\ 1997, \aaps, 124,
\bibitem[Metcalfe et al.(2015)]{met15} Metcalfe, T.~S., Creevey, O.~L., \& Davies, G.~R.\ 2015, \apjl, 811, L37 
\bibitem[Mishenina et al.(2016)]{mis16} Mishenina, T., Kovtyukh, V., Soubiran, C., \& Adibekyan, V.~Z.\ 2016, \mnras, 462, 1563 
\bibitem[Monroe et al.(2013)]{mon13} Monroe, T.~R., Mel{\'e}ndez, J., Ram{\'{\i}}rez, I., et al.\ 2013, \apjl, 774, L32 
\bibitem[Montes et al.(2001)]{montes01} Montes, D., L{\'o}pez-Santiago, J., Fern{\'a}ndez-Figueroa, M.~J., \& G{\'a}lvez, M.~C.\ 2001, \aap, 379, 976
\bibitem[Nissen(2015)]{nis15} Nissen, P.~E.\ 2015, \aap, 579, A52
\bibitem[Nissen et al.(2017)]{nis17} Nissen, P.~E., Silva Aguirre, V., Christensen-Dalsgaard, J., et al.\ 2017, \aap, 608, A112 
\bibitem[Nissen \& Gustafsson(2018)]{nis18} Nissen, P.~E., \& Gustafsson, B.\ 2018, \aapr, 26, 6
\bibitem[Noguchi et al.(2002)]{nog02} Noguchi, K., Aoki, W., Kawanomoto, S., et al.\ 2002, \pasj, 54, 855 
\bibitem[Oh et al.(2018)]{oh18} Oh, S., Price-Whelan, A.~M., Brewer, J.~M., et al.\ 2018, \apj, 854, 138 
\bibitem[{\"O}nehag et al.(2014)]{one14} {\"O}nehag, A., Gustafsson, B., \& Korn, A.\ 2014, \aap, 562, A102
\bibitem[Pasquini \& Pallavicini(1991)]{pasquini91} Pasquini, L., \& Pallavicini, R.\ 1991, \aap, 251, 199
\bibitem[Patience et al.(2002)]{pat02} Patience, J., White, R.~J., Ghez, A.~M., et al.\ 2002, \apj, 581, 654 
\bibitem[Petrovich \& Mu{\~n}oz(2017)]{pet17} Petrovich, C., \& Mu{\~n}oz, D.~J.\ 2017, \apj, 834, 116 
\bibitem[Pl{\'a}valov{\'a} \& Solovaya(2013)]{pla13} Pl{\'a}valov{\'a}, E., \& Solovaya, N.~A.\ 2013, \aj, 146, 108
\bibitem[Ram{\'{\i}}rez et al.(2009)]{ram09} Ram{\'{\i}}rez, I., Mel{\'e}ndez, J., \& Asplund, M.\ 2009, \aap, 508, L17
\bibitem[Ram{\'{\i}}rez et al.(2010)]{ram10} Ram{\'{\i}}rez, I., Asplund, M., Baumann, P., Mel{\'e}ndez, J., \& Bensby, T.\ 2010, \aap, 521, A33
\bibitem[Ram{\'{\i}}rez et al.(2011)]{ram11} Ram{\'{\i}}rez, I., Mel{\'e}ndez, J., Cornejo, D., Roederer, I.~U., \& Fish, J.~R.\ 2011, \apj, 740, 76 
\bibitem[Ram{\'{\i}}rez et al.(2013)]{ram13} Ram{\'{\i}}rez, I., Allende Prieto, C., \& Lambert, D.~L.\ 2013, \apj, 764, 78 
\bibitem[Ram{\'{\i}}rez et al.(2014)]{ram14} Ram{\'{\i}}rez, I., Mel{\'e}ndez, J., Bean, J., et al.\ 2014, \aap, 572, A48 
\bibitem[Ram{\'{\i}}rez et al.(2015)]{ram15} Ram{\'{\i}}rez, I., Khanal, S., Aleo, P., et al.\ 2015, \apj, 808, 13
\bibitem[van Saders et al.(2016)]{san16} van Saders, J.~L., Ceillier, T., Metcalfe, T.~S., et al.\ 2016, \nat, 529, 181
\bibitem[Saffe et al.(2015)]{saf15} Saffe, C., Flores, M., \& Buccino, A.\ 2015, \aap, 582, A17
\bibitem[Saffe et al.(2016)]{saf16} Saffe, C., Flores, M., Jaque Arancibia, M., Buccino, A., \& Jofr{\'e}, E.\ 2016, \aap, 588, A81
\bibitem[Saffe et al.(2017)]{saf17} Saffe, C., Jofr{\'e}, E., Martioli, E., et al.\ 2017, \aap, 604, L4 
\bibitem[Saffe et al.(2019)]{saf19} Saffe, C., Jofr{\'e}, E., Miquelarena, P., et al.\ 2019, \aap, 625, A39
\bibitem[dos Santos et al.(2016)]{leo16} dos Santos, L.~A., Mel{\'e}ndez, J., do Nascimento, J.-D., et al.\ 2016, \aap, 592, A156 
\bibitem[Sandquist et al.(2002)]{san02} Sandquist, E.~L., Dokter, J.~J., Lin, D.~N.~C., \& Mardling, R.~A.\ 2002, \apj, 572, 1012
\bibitem[Schuler et al.(2011)]{sch11} Schuler, S.~C., Flateau, D., Cunha, K., King, J.~R., Ghezzi, L., \& Smith, V.~V.\ 2011, \apj, 732, 55 
\bibitem[da Silva et al.(2012)]{sil12} da Silva, R., Porto de Mello, G.~F., Milone, A.~C., et al.\ 2012, \aap, 542, A84 
\bibitem[Sneden(1973)]{sne73} Sneden, C.~A.\ 1973, Ph.D.~Thesis, 
\bibitem[Sousa(2014)]{sou14} Sousa, S.~G.\ 2014, Determination of Atmospheric Parameters of B-, A-, F- and G-Type Stars.~Series: GeoPlanet: Earth and Planetary Sciences, ISBN: <ISBN>978-3-319-06955-5</ISBN>.~Springer International Publishing (Cham), Edited by Ewa Niemczura, Barry Smalley and Wojtek Pych, pp.~297-310, 297 
\bibitem[Sousa et al.(2015)]{sou15} Sousa, S.~G., Santos, N.~C., Adibekyan, V., Delgado-Mena, E., \& Israelian, G.\ 2015, \aap, 577, A67 
\bibitem[Spina et al.(2015)]{spi15} Spina, L., Palla, F., Randich, S., et al.\ 2015, \aap, 582, L6\bibitem[Stetson \& Pancino(2008)]{ste08} Stetson, P.~B., \& Pancino, E.\ 2008, \pasp, 120, 1332 
\bibitem[Spina et al.(2016)]{spi16} Spina, L., Mel{\'e}ndez, J., \& Ram{\'{\i}}rez, I.\ 2016, \aap, 585, A152
\bibitem[Spina et al.(2018)]{spi18} Spina, L., Mel{\'e}ndez, J., Karakas, A.~I., et al.\ 2018, \mnras, 474, 2580
\bibitem[Takeda et al.(2002)]{tak02} Takeda, Y., Zhao, G., Chen, Y.-Q., Qiu, H.-M., \& Takada-Hidai, M.\ 2002, \pasj, 54, 275
\bibitem[Takeda et al.(2011)]{tak11} Takeda, Y., Tajitsu, A., Honda, S., et al.\ 2011, \pasj, 63, 697
\bibitem[Th{\'e}ado \& Vauclair(2012)]{the12} Th{\'e}ado, S., \& Vauclair, S.\ 2012, \apj, 744, 123 
\bibitem[Teske et al.(2015)]{tes15} Teske, J.~K., Ghezzi, L., Cunha, K., et al.\ 2015, \apjl, 801, L10
\bibitem[Teske et al.(2016)]{tes16} Teske, J.~K., Khanal, S., \& Ram{\'{\i}}rez, I.\ 2016, \apj, 819, 19 
\bibitem[Teske et al.(2016b)]{tes16b} Teske, J.~K., Shectman, S.~A., Vogt, S.~S., et al.\ 2016, \aj, 152, 167
\bibitem[Tucci Maia et al.(2014)]{tuc14} Tucci Maia, M., Mel{\'e}ndez, J., \& Ram{\'{\i}}rez, I.\ 2014, \apjl, 790, L2
\bibitem[Tucci Maia et al.(2016)]{tuc16} Tucci Maia, M., Ram{\'{\i}}rez, I., Mel{\'e}ndez, J., et al.\ 2016, \aap, 590, A32 
\bibitem[Tucci Maia et al.(2015)]{tuc15} Tucci Maia, M., Mel{\'e}ndez, J., Castro, M., et al.\ 2015, \aap, 576, L10
\bibitem[Turner et al.(2001)]{tur01} Turner, N.~H., ten Brummelaar, T.~A., McAlister, H.~A., et al.\ 2001, \aj, 121, 3254
\bibitem[Wittenmyer et al.(2007)]{wit07} Wittenmyer, R.~A., Endl, M., \& Cochran, W.~D.\ 2007, \apj, 654, 625 
\bibitem[Yi et al.(2001)]{yi01} Yi, S., Demarque, P., Kim, Y.-C., et al.\ 2001, \apjs, 136, 417
\end{thebibliography}
\end{document}